\documentclass[prd,amsmath,amssymb,superscriptaddress,nofootinbib,twocolumn]{revtex4}
\usepackage{amsfonts}
\usepackage{multirow}
\usepackage{makecell}
\usepackage{mathrsfs}
\usepackage{graphicx}
\usepackage{amsmath}
\usepackage{amssymb}
\usepackage{bm}
\usepackage{bbm}
\usepackage{color}
\usepackage{ulem}
\usepackage{array}
\usepackage{diagbox}
\usepackage{float}
\allowdisplaybreaks[4]

\newcommand{\nn}{\nonumber}

\newcommand{\beq}{\begin{equation}}
\newcommand{\eeq}{\end{equation}}
\newcommand{\bqa}{\begin{eqnarray}}
\newcommand{\eqa}{\end{eqnarray}}

\newcommand{\bseq}{\begin{subequations}}
\newcommand{\eseq}{\end{subequations}}

\makeatletter

\begin{document}

\title{Light quark mass dependence of nucleon mass to two-loop order
}
\author{Long-Bin Chen~\footnote{chenlb@gzhu.edu.cn}}
\affiliation{ School of Physics and Materials Science, Guangzhou University, Guangzhou 510006, China \vspace{0.2cm}}
\author{Siwei Hu~\footnote{husw@ihep.ac.cn}}
\affiliation{Institute of High Energy Physics, Chinese Academy of Sciences, Beijing 100049, China\vspace{0.2cm}}
\affiliation{School of Physical Sciences, University of Chinese Academy of Sciences, Beijing 100049, China\vspace{0.2cm}}
\author{Yu Jia~\footnote{jiay@ihep.ac.cn}}
\affiliation{Institute of High Energy Physics, Chinese Academy of Sciences, Beijing 100049, China\vspace{0.2cm}}
\affiliation{School of Physical Sciences,
University of Chinese Academy of Sciences, Beijing 100049, China\vspace{0.2cm}}
\author{Zhewen Mo\footnote{mozw@itp.ac.cn}}
\affiliation{CAS Key Laboratory of Theoretical Physics, Institute of Theoretical Physics,
	Chinese Academy of Sciences, Beijing 100190, China\vspace{0.2cm}}
\affiliation{Institute of High Energy Physics, Chinese Academy of
	Sciences, Beijing 100049, China\vspace{0.2cm}}

\date{\today}

\begin{abstract}
We investigate the nucleon self energy through the sixth chiral order in the covariant $SU(2)$ chiral perturbation theory ($\chi$PT)
in the single baryon sector.  The validity of the extended on-mass-shell (EOMS) renormalization scheme is explicitly verified to two-loop order,
manifested by the miraculous cancellation of all nonlocal divergences and power-counting-breaking (PCB) terms that are nonanalytic in pion mass.
Using the $\sigma_{\pi N}$ term determined from the latest lattice simulation to constrain some unknown higher-order low energy constants (LECs),
we predict the nucleon mass in the chiral limit to be $856.6\pm 1.7$ MeV.
It is found that the EOMS scheme exhibits quite satisfactory convergence behavior through ${\cal O}(q^6)$ around physical point.
We also predict the pion mass dependence of the nucleon mass to the accuracy of ${\cal O}(q^6)$,
which is in satisfactory agreement with the recent lattice results over a wide range of pion mass.
\end{abstract}

\maketitle

\noindent{\color{blue}\it Introduction.} As the highly relativistic bound states formed by light quarks and gluons,
proton and neutron comprise the basic building blocks in our visible universe.
Understanding various facets of nucleons from quantum chromodynamics (QCD), the underlying theory of strong interaction,
is one of the most outstanding challenges faced by the contemporary particle and nuclear physics.
For example, the central goal of the existing and forthcoming $ep$ facilities such as
{\tt JLab}, {\tt EIC} and {\tt EicC} is to unravel the internal structure of proton to an unprecedented accuracy.

To gain a deeper insight into QCD, one is not necessarily confined within the realistic world.
It is of fundamental interest to ascertain how the nucleon mass varies with the light quark mass,
or equivalently, the pion mass.
Two first-principle approaches, {\it i.e.}, the lattice Monte Carlo simulations and chiral perturbation theory ($\chi$PT) are tailored to address such a quest.
There have arose intensive investigations from the purely numerical lattice approach over the past two decades~\cite{JLQCD:2002zto,Procura:2003ig,Procura:2006bj,Alexandrou:2009qu,Alexandrou:2014sha,RBC:2014ntl,Alexandrou:2017xwd,Yang:2018nqn,RQCD:2022xux}
(for a recent review, see {\tt FLAG 2021} report~\cite{FlavourLatticeAveragingGroupFLAG:2021npn}).
Some while ago, the so-called ruler approximation, {\it i.e.}, a simple functional form of 800 MeV plus pion mass,
was found to roughly reproduce the lattice results of the pion mass dependence of the nucleon mass
over a wide range~\cite{Walker-Loud:2008rui,Walker-Loud:2014iea}. However, bearing a wrong leading $M_\pi$ dependence and in absence of the chiral logarithms,
this simple parameterization diametrically contradicts the prediction made from the $\chi$PT.
Needless to say, a critical comparison between the analytical $\chi$PT prediction and lattice calculation is invaluable to clarify the situation.
Moreover, the expense of lattice calculation becomes quickly unaffordable for a pion mass much lighter
than the physical one, but $\chi$PT serves the only analytic, model-independent approach
to explore this regime, especially for predicting the nucleon mass in the chiral limit.

Chiral perturbation theory is the time-honored low-energy effective field theory of QCD,
with the pions~\cite{Weinberg:1978kz,Gasser:1983yg} and nucleons~\cite{Gasser:1987rb} being the active degrees of freedom.
Since the nucleon mass does not vanish in the chiral limit, which numerically coincides with the chiral symmetry breaking scale $4\pi F_\pi$,
the single-baryon sector of $\chi$PT is plagued with the notorious power-counting (PC) problem~\cite{Gasser:1987rb}.
Three influential schemes have been proposed to overcome the PC problem, exemplified by the heavy baryon (HB) formalism~\cite{Jenkins:1990jv},
the infrared (IR) regularization~\cite{Becher:1999he} and the extended-on-mass-shell (EOMS) renormalization scheme~\cite{Gegelia:1999gf, Fuchs:2003qc}.
The chiral expansion of nucleon mass has been computed to ${\cal O}(q^4)$ in the EOMS scheme~\cite{Fuchs:2003qc},
to ${\cal O}(q^5)$ in HB approach~\cite{McGovern:1998tm}, and to the ${\cal O}(q^6)$ accuracy in the IR scheme~\cite{Schindler:2006ha, Schindler:2007dr}.

At practical level, it turns out that the Lorentz-invariant EOMS scheme exhibits faster convergence and better analyticity behavior
with respect to two other schemes in numerous situations, taking  $\pi N$ scattering~\cite{Fettes:2000xg,Alarcon:2012kn,Chen:2012nx, Siemens:2016hdi}, the magnetic moments of octet baryons~\cite{Durand:1997ya,Geng:2008mf}, $\overline{K} N$ scattering~\cite{Lu:2022hwm} and chiral nucleon-nucleon interaction~\cite{Lu:2021gsb}
as examples (for a review on phenomenological success of the EOMS scheme applied in various situations, see \cite{Geng:2013xn}).

It is the central aim of this work to investigate the nucleon self energy through ${\cal O}(q^6)$ in EOMS scheme.
This work entails analytically computing the Lorentz-invariant two-loop integrals with two distinct mass scales,
which is technically more involved than its HB and IR counterparts~\cite{McGovern:1998tm,Schindler:2006ha, Schindler:2007dr}.
We explicitly verify the self consistency of the EOMS scheme at two-loop order, which turns out to be highly nontrivial.
Once deducing the chiral expansion of the nucleon mass to the ${\cal O}(q^6)$ accuracy,
we then present the most precise numerical prediction for the nucleon mass in the chiral limit.
Finally we make a critical comparison between the state-of-the-art $\chi$PT predictions
and the recent lattice results for the pion mass dependence of the nucleon mass.

\vspace{0.2 cm}
\noindent{\color{blue}\it Outline of calculation.}  Our starting point is
the $SU(2)$ $\chi$PT lagrangian for pion and nucleons in the isospin symmetric limit:
\beq
\label{Lag}
{\cal L}_{\chi{\rm PT}} = {\cal L}_{\pi\pi}^{(2)} + {\cal L}_{\pi\pi}^{(4)} + \sum_{i=1}^6 {\cal L}^{(i)}_{{\pi}N}+\cdots.
\eeq
The purely mesonic sector through ${\cal O}(q^4)$ is well-known~\cite{Gasser:1987rb}.
The lowest-order single nucleon chiral Lagrangian reads~\cite{Gasser:1987rb}
\begin{equation}
     \mathcal{L}_{\pi N}^{(1)} = \bar{\Psi} \left( i{\not\!\partial} -m +
     \frac{{\tt g_A}}{2F} \gamma_5 \not\!\partial \vec{\pi}\cdot \vec{\tau} \right) \Psi + \cdots,
\end{equation}
where $\Psi=(p\;\:n)^T$ and $\vec{\pi}$ represent the Dirac fields for nucleon doublet and the fields for the pion iso-triplet, $\vec{\tau}$
denotes the Pauli matrix in the isospin space. $m$, $F$ and ${\tt g_A}$ denote the nucleon mass, pion decay constant and the axial-vector coupling in the chiral limit.
We follow the conventions of \cite{Fettes:2000gb} for the nucleon chiral lagrangian through ${\cal O}(q^4)$,
with $c_i$, $d_i$, $e_i$ representing the low energy constants (LECs) affiliated with the chiral order $2$, $3$, $4$, respectively.
Very recently $\mathcal{O}\left(q^5\right)$ baryon chiral lagrangian has been constructed with the aid of modern on-shell amplitude method~\cite{Song:2024fae},
which nevertheless is not needed in this work. Although the precise texture of ${\cal L}^{(6)}_{{\pi}N}$ is not yet available,
its net contribution to nucleon self energy can be summarized by a tree-level contact term $\hat{g}_1 M^6$,
with $\hat{g}_1$ signalling an unknown LEC~\cite{McGovern:1998tm, Schindler:2006ha, Schindler:2007dr}.

The physical nucleon mass can be determined by locating the pole position of the nucleon's propagator:
\beq
\left\{{\not\!p} - m_{\tt B} -\Sigma({\not\! p}, m_{\tt B})\right\} {\big \vert}_{\,{\not\! p}=m_N} = 0,
\label{Pole:position:physical:nucleon:mass}
\eeq
where $m_{\tt B}$ denotes the bare nucleon mass and $\Sigma$ signifies the contribution of all the one-particle irreducible (1PI) self-energy diagrams.
We find it convenient in this work to utilize the unrenormalized perturbation theory and start with the bare fields, masses and couplings.

Eq.~\eqref{Pole:position:physical:nucleon:mass} will be solved in an iterative manner.
Schematically, the physical nucleon mass can be expanded as follows:
\beq
m_N = m_{\tt B} + \Sigma_{\tt c} + \hbar\Sigma^{(1)} + \hbar^2\Sigma^{(2)} + \mathcal{O}(\hbar^3),
\label{mN:expanded:in:bare:quantities}
\eeq
where the contact term $\Sigma_{\tt c} = -4c_1M^2 -2(8e_{38}+e_{115}+e_{116})M^4+\hat{g}_1M^6$  absorbs a series of tree-level contributions to the
self energy from various chiral orders, where $M$ designates the pion mass determined via the Gell-Mann-Oakes-Renner relation.
Here $\hbar$ is explicitly inserted as a bookkeeping device to trace the loop order affiliated with the self energy contributions.

\begin{figure}[ht]
\centering
\includegraphics[width=0.5\textwidth]{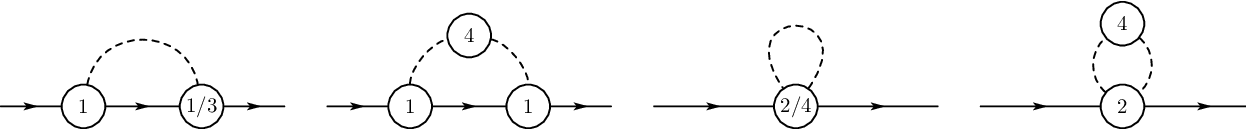}
\caption{One-loop Feynman diagrams contributing to nucleon self energy through ${\cal O}(q^6)$.
The solid line signifies the nucleon, while the dashed curve represents the pion.
The number inside each blob denotes the chiral order of the respective vertex.
For simplicity, we have suppressed those diagrams with the $\Sigma_{\tt c}$ vertex inserted onto the nucleon propagators.}
\label{Feyn:Diagrams:one:loop}
\end{figure}

\begin{figure}[ht]
    \centering
    \includegraphics[width=0.5\textwidth]{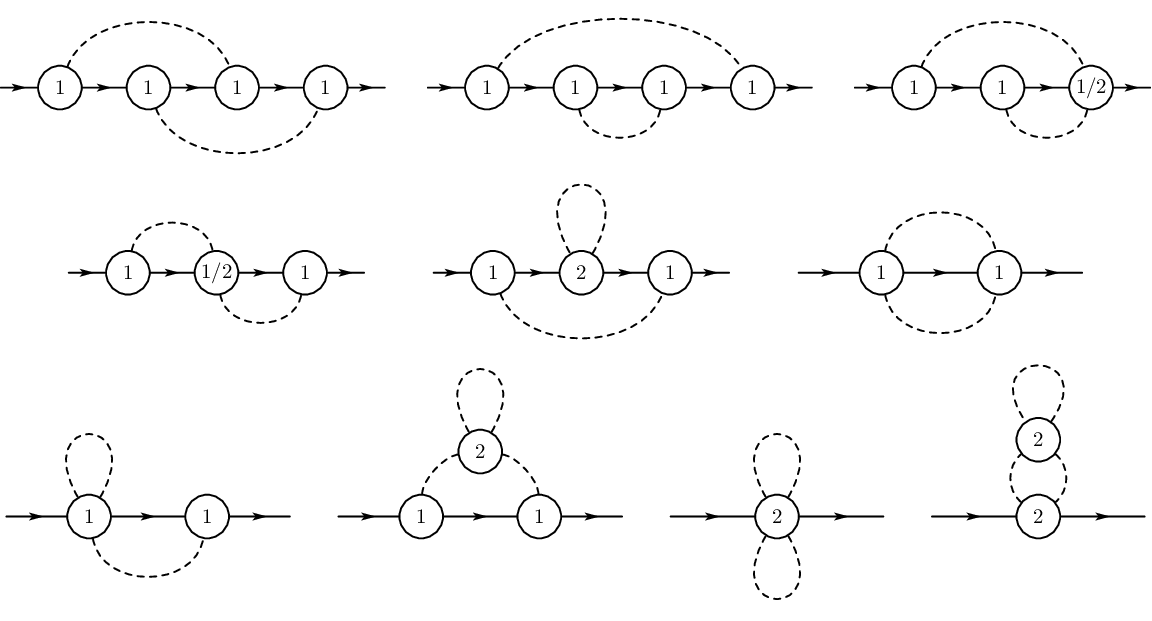}
    \caption{Two-loop diagrams contributing to nucleon self energy through ${\cal O}(q^6)$. The mirror diagrams have been
    neglected for simplicity.  We have also suppressed those diagrams with the
    $\Sigma_{\tt c}$ vertex inserted onto the nucleon propagators.}
\label{Feyn:Diagrams:two:loop}
\end{figure}

The one and two-loop self energy diagrams through ${\cal O}(q^6)$ are depicted in Fig.~\ref{Feyn:Diagrams:one:loop} and Fig.~\ref{Feyn:Diagrams:two:loop}.
The nominal chiral order of each diagram is $D = d N_L + \sum_i k_i - I_N - I_\pi$, where $d$ is the spacetime dimension,
$N_L$ represents the loop order, $I_N$ and $I_\pi$ are the numbers of nucleon and pion propagators,
$k_i$ is the chiral order of the $i$-th vertex.
We adopt dimensional regularization with $d=4-2\epsilon$ to regularize the UV divergences arising
from loop integrations.

Note that those self energy diagrams with the $\Sigma_{\tt c}$ vertex inserted onto the nucleon propagators need also be considered.
Since $\Sigma_{\tt c}$ is a constant rather than function of $p^2$,  one can readily resum their contributions,
which amounts to shifting the nucleon mass $m_{\tt B}$ to $\tilde{m}\equiv m_{\tt B}+\Sigma_{\tt c}$ in the nucleon propagator.
We use the symbol $\tilde{\Sigma}^{(1,2)}(m_N, \tilde{m})$  to signify the one and two-loop self energy contributions
with the modified nucleon propagator and without the insertion of the $\Sigma_{\tt c}$ vertex~\cite{Schindler:2006ha, Schindler:2007dr}.

Since $m_N\neq \tilde{m}$, one needs to consider the off-shell two-point Feynman integrals.
It is straightforward to calculate the one-loop self energy rigorously, followed by
replacing all the occurrences of bare parameters in the resulting analytic
expression of $\tilde{\Sigma}^{(1)}$ with the renormalized ones plus the counterterms
dictated in the EOMS scheme:
\begin{subequations}
\bqa
&& l_{i, {\tt B}} = l_{i, {\tt R}} + \delta_{\text{div}}l_{i},
\\
&& x_{j, {\tt B}} = x_{j, \text{R}} + \delta_{\text{div}} x_j + \delta_{\text{eoms}} x_j,
\\
&& m_{\tt B} = m + \delta_{\text{div}} m + \delta_{\text{eoms}} m,
\\
&& {\tt g_{A,B}} = {\tt g_A} + \delta_{\text{div}} {\tt g_A} + \delta_{\text{eoms}} {\tt g_A},
\eqa
\label{Bare:parameters:renormalized}
\end{subequations}
where $l_i$ signify the LECs in purely mesonic sector, and $x_j \in \left\{c_j, d_j, e_j\right\}$ denote the LECs in the baryon sector.
The divergent counterterms $\delta_{\text{div}}$ are determined under the $\widetilde{\text{MS}}$ (also called $\overline{\rm MS}-1$)
scheme. To facilitate the $\widetilde{\text{MS}}$ subtraction in the multiloop case,
one simply shifts the 't Hooft unit mass, so that the prefactor $\mu^{2\epsilon}$ affiliated with each loop integral
is modified into $\mu^{2\epsilon}e^{\epsilon(\gamma_E - 1)}/(4\pi)^{2\epsilon}$,
and the $\delta_{\text{div}}$  can be determined from the powers of $1/\epsilon$ poles in the resulting loop integrations~\cite{Schindler:2007dr}.
After subtracting the UV poles, those finite power-counting breaking (PCB) terms need be further subtracted in the EOMS scheme.
The corresponding finite PCB counterterms are referred to as $\delta_{\text{eoms}}$ in \eqref{Bare:parameters:renormalized}.
Note all $\delta_{\text{div}}$ and $\delta_{\text{eoms}}$ begins with ${\cal O}(\hbar)$.

Here arises a technical complication. Since the one-loop self energy diagrams are generally logarithmically UV divergent,
one has to determine the $\delta_{\text{eoms}}$ appearing in Fig.~\ref{Feyn:Diagrams:one:loop} to $\mathcal{O}(\epsilon)$,
which could result in a  finite (PCB) contribution in one-loop counterterm diagram.
By reinvestigating the $\pi N$ scattering amplitude at one-loop order and to ${\cal O}(\epsilon)$, we are able to
determine all the required $\delta_{\text{eoms}} \{m, {\tt g_A}, c_j, d_j, e_j\}$  to ${\cal O}(\epsilon)$ accuracy.
For reader's convenience, we have enumerated all the encountered counterterms in EOMS scheme
in Appendix~\ref{counterterms:EOMS}.

On the other hand, calculating the off-shell two-loop self energy turns out to be a formidable task,
in which elliptic functions would inevitably arise in our case.
Fortunately, it is legitimate to replace $m_N$ in $\tilde{\Sigma}^{(2)}(m_N, \tilde{m})$  by $\tilde{m}$,
since the neglected $\hbar \Sigma^{(1)}$ term in \eqref{mN:expanded:in:bare:quantities} would yield an overall ${\cal O}(\hbar^3)$ contribution,
which is beyond the intended accuracy of this work. Hence we can simply work with $\tilde{\Sigma}^{(2)}(\tilde{m}, \tilde{m})$,
consequently manipulating on-shell two-loop integrals becomes much more manageable.

In actual calculation, we use some in-house codes to generate Feynman amplitudes and
manipulate the Lorentz/Dirac/isospin algebra. We employ the package {\tt FIRE}~\cite{Smirnov:2019qkx}
for integration-by-part reduction.
For the two-loop self energy, we end up with 13 master integrals (MIs).
The differential equation method~\cite{Kotikov:1990kg,Kotikov:1991pm,Henn:2013pwa} has been employed to work out
the analytic expressions of these MIs, which are expressed in terms of the Goncharov polylogarithms~\cite{Goncharov:1998kja}.
The correctness of these analytic expressions is numerically checked by utilizing the package {\tt AMFlow}~\cite{Liu:2022chg}.
Technical details about computing these MIs will be presented elsewhere.
For readers' convenience, in Appendix~\ref{analytical:expression:two:loop:Sigma}
we present the expanded expression of $\tilde{\Sigma}^{(2)}$ up to ${\cal O}(q^6)$.

Replacing the bare parameters in $\tilde{\Sigma}^{(1,2)}$ with the renormalized ones in accordance with \eqref{Bare:parameters:renormalized},
conducting the chiral expansion~\footnote{It might be worth emphasizing that, since $\tilde{m}$ contains
contributions from different chiral order, one should expand $\tilde{m}$ in $\tilde{\Sigma}^{(1,2)}(\tilde{m}, \tilde{m})$
around $m$ rather than $\tilde{m}_{\tt R} \equiv m + \Sigma_{{\tt c}, {\tt R}}$,
otherwise some higher-order power-counting preserving (PCP) terms in $\tilde{m}$ might be erroneously subtracted
together with the PCB terms.},
substituting them into \eqref{mN:expanded:in:bare:quantities},
and truncating the expressions to order  $q^6$, $\hbar^2$ and $\epsilon^0$,  we find that
intriguingly all the nonlocal divergences and PCB terms that are non-analytic in $M_\pi$ in two-loop diagrams are
exactly cancelled by the one-loop counterterm diagrams entailing $\delta_{\text{div}}$ and $\delta_{\text{eoms}}$.
This miraculous cancelation provides a highly nontrivial consistency check of EOMS scheme at two-loop level.
The remaining divergences and PCB terms only arise at even chiral order and are analytic in $\ln M_\pi$,
which can be readily absorbed in the tree-level ${\cal O}(\hbar^2)$ counterterms~\cite{Schindler:2003je}.

\vspace{0.2 cm}
\noindent{\color{blue}\it Chiral expansion of nucleon mass and $\sigma_{\pi N}$ to ${\cal O}(q^6)$.}
Through $\mathcal{O}\left(q^6\right)$, the physical nucleon mass is subject to
the following chiral expansion:
\bqa
m_N &=&  m + k_1 M^2 + k_2 M^3 + k_3 M^4 \ln\frac{M}{\mu} + k_4 M^4
\nn\\
&+& k_5 M^5 \ln\frac{M}{\mu} + k_6 M^5
\label{mN:chiral:expansion:q6}
\\
&+& k_7 M^6 \ln^2\frac{M}{\mu} + k_8 M^6 \ln\frac{M}{\mu}+ k_9 M^6,
\nn
\eqa
where $\mu$ represents the renormalization scale in the EOMS scheme.
The full expression of \eqref{mN:chiral:expansion:q6} is rather lengthy, and we are content with
showing the expressions of $k_i$ ($i=1,\cdots 9$) with $\mu=m$ for simplicity.
First we reproduce the well-known expressions of $k_1$ through $k_4$ in EOMS scheme,
originally given in \cite{Fuchs:2003qc}~\footnote{We emphasize that all the parameters in \eqref{mN:chiral:expansion:q6} are finite.
Without causing confusion, we have omitted the subscript ``{\tt R}'' in all LECs, which are evaluated at $\mu=m$.}:
\begin{subequations}
\bqa
&& k_1 = -4c_1,
\\
&& k_2  = -\frac{3 {\tt g_A}^2}{32F^2\pi},
\\
& & k_3 = -\frac{3 {\tt g_A}^2}{32 \pi ^2 F^2 m} + \frac{3 \left(8 c_1-c_2-4 c_3\right)}{32 \pi ^2 F^2},
\\
& &  k_4 = \frac{3 {\tt g_A}^2 \left(1+4 c_1 m\right)}{32 \pi ^2 F^2 m}+\frac{3 c_2}{128 \pi ^2 F^2} - \hat{e}_1.
\eqa
\label{expressions:k1:to:k4}
\end{subequations}

The major new result of this work is the knowledge of $k_5$ through $k_9$ in EOMS scheme~\footnote{We notice that, in a recent proceeding article~\cite{Conrad:2024phd}, Conrad, Gasparyan and Epelbaum have reported the analytical results of the renormalized nucleon self-energy through $q^6$ in EOMS scheme. Rather than compute the two-loop integrals rigourously then conduct the chiral expansion, the authors of \cite{Conrad:2024phd} apply the strategy of region~\cite{Beneke:1997zp} to
compute the two-loop diagrams. Since the renormalized self energy in \cite{Conrad:2024phd} is expressed in terms of $m_N$ instead of $m$,
a direct comparison between theirs and our \eqref{mN:chiral:expansion:q6} is not straightforward.
Nevertheless,  factors such as $\zeta_3$ and $\pi^2 \ln 2$ in \eqref{expression:k9},
both of which have degree three of transcendentality, seem to be absent in their expression of two-loop
renormalized self energy.}:
\begin{subequations}
\bqa
&& k_5 = {3 {\tt g_A}^2 \over 1024 \pi^3 F^4} \left(16{\tt g_A}^2-3\right),
\label{expression:k5}
\\
&& k_6 = \frac{17 {\tt g_A}^4}{512 \pi ^3 F^4} -\frac{3 \hat{d}_{16} {\tt g_A}}{8 \pi  F^2}
\\
&&  + \frac{3{\tt g_A}^2 \big[\pi ^2 F^2+2m^2 + 8m^2\pi ^2 \left(2 l_4-3 l_3\right)\big]}{256\pi ^3 F^4 m^2},
\nn\\
 &&  k_7 = - {3\over 256 \pi ^4 F^4m} \left[{\tt g_A}^2 - m(6 c_1 - c_2 - 4 c_3)\right],
\label{expression:k7}
\\
 && k_8 = \frac{-3}{8 \pi^2 F^2 m^2}\left[
            c_1 {\tt g_A}^2 + m \left(\hat{d}_{16} {\tt g_A}-2 c_1 c_2\right) \right.
 \nn\\
&& \left.
+\left(8 e_{14}+2 e_{15}+e_{16}+2
            \hat{e}_{20}+4 \hat{e}_{36}+4 \hat{e}_{38}\right) m^2
        \right]
\nn\\
 && +\frac{1}{3072F^4m^2\pi^4}\left[
 114 {\tt g_A}^4+\left(51-576\pi^2 \left(2 l_3-l_4\right)\right) {\tt g_A}^2
            \right.
\nn\\
& &+36 c_1 m \left(-6-25 {\tt g_A}^4+10 {\tt g_A}^2+128 \pi ^2 \left(l_3-l_4\right)\right)
\\
&& -c_2 m \left(576 \pi ^2 \left(2 l_3-l_4\right)+23 {\tt g_A}^2\right)
\nn\\
& & + 4 c_3 m \left(-9+41 {\tt g_A}^2-576 \pi ^2 \left(2 l_3-l_4\right)\right)
\nn\\
&& \left. -4 c_4 m \left(21+13 {\tt g_A}^2\right) -18\right],
\nn\\
&& k_9 = \hat{g}_1 + \frac{c_1 \left(768 l_3+7\right)-4 \left(96 c_3 l_3+12 c_2 l_4+c_4\right)}{1024 \pi^2   F^4}
\nn\\
&& +5\frac{ \left(225 c_1+4 c_4\right)m+6}{12288 \pi ^4 F^4 m}  +\frac{{\tt g_A}^2 \left(704 c_1^3+3 \hat{e}_1\right)}{32\pi ^2 F^2}
\nn\\
&& +\frac{{\tt g_A}^2 \left(144 c_1 \left(128 l_3-128 l_4-3\right)+83 c_2+512 c_3-928c_4\right)}{24576 \pi ^2 F^4}
\nn\\
&& -\frac{\left(6210 c_1-1147 c_2+616 c_3-948 c_4\right) {\tt g_A}^2}{12288 \pi ^4 F^4}
\label{expression:k9}
\\
& &  - { {\tt g_A}^2\over 128 \pi ^2 F^2 m^3}  + \frac{{\tt g_A}^2\left(-1+18 l_3-12 l_4\right)}{64 \pi ^2 F^4m}
\nn\\
&& + \frac{ c_1 {\tt g_A}^4 (96\pi^2\ln 2-144 \zeta_3-164\pi^2+1779) }{4096 \pi ^4 F^4}
\nn\\
& &+\frac{{\tt g_A}^2(18-(41\pi^2+145) {\tt g_A}^2)}{2048 \pi ^4 F^4 m} +
   \frac{6 e_{15}+5 e_{16}+6 \hat{e}_{20}}{32 \pi ^2 F^2}
\nn\\
& & +\frac{-24 c_1^2
   {\tt g_A}^2+6 \hat{d}_{16} {\tt g_A}-3 c_1 c_2}{16 \pi ^2 F^2 m} +\frac{3 c_1 \hat{d}_{16} {\tt g_A}}{2 \pi ^2 F^2},\nn
\eqa
\label{Expressions:k5:to:k9}
\end{subequations}
with the hatted LECs defined by
\begin{subequations}
\bqa
&& \hat{d}_{16} = 2d_{16} - d_{18},
\\
&&\hat{e}_{1~} = 2\left(8e_{38}+e_{115}+e_{116}\right),
\\
&&\hat{e}_{20} = e_{20} + e_{35},
\\
&&\hat{e}_{36} = 2e_{19} - e_{22} - e_{36},
\\
&&\hat{e}_{38} = e_{22} - 4e_{38}.
\label{eq:k1k9}
\eqa
\end{subequations}
The chiral logarithm is of special interest, which has captured the infrared non-analytic behavior of QCD.
The chiral logarithm at order $q^5$,  $k_5$, only depends on ${\tt g_A}$, whereas the leading double logarithm at order $q^6$,
$k_7$, depends only on ${\tt g_A}$ and the ${\cal O}(q^2)$ LECs.
This pattern is dictated by renormalization group invariance~\cite{Bijnens:2014ila}.
It is also worth emphasizing that, the expression of $k_5$, \eqref{expression:k5}, is identical with its HB~\cite{McGovern:1998tm} and
IR counterparts~\cite{Schindler:2006ha, Schindler:2007dr},
and the expression of $k_7$, \eqref{expression:k7}, is identical with its IR counterpart~\cite{Schindler:2006ha, Schindler:2007dr}.
In fact, the leading chiral logarithm at each chiral order must be scheme-independent,
since they cannot be modified by the redefinitions of LECs~\cite{Schindler:2006ha,Schindler:2007dr,Bijnens:2014ila}.

An important observable closely related to nucleon mass is the $\pi N$ sigma term, which quantifies the contribution to the nucleon mass
from the masses of $u$ and $d$ quarks~\cite{Ji:1994av}, and is of also great interest in direct detection of dark matter particles (for a recent review on the phenomenological status of the $\sigma_{\pi N}$ term,
see Refs.~\cite{Hoferichter:2015hva,Ren:2017fbv}.
As a bonus, one can derive from \eqref{mN:chiral:expansion:q6} the chiral expansion of the $\pi N$ sigma term
with the aid of  Hellmann-Feynman theorem:
\bqa
&& \sigma_{\pi N} = M^2 \frac{\partial m_N}{\partial M^2}
\nn\\
& &=k_1M^2 + \frac{3}{2}k_2M^3+ 2k_3M^4\ln \frac{M}{\mu} + \frac{k_3+4k_4}{2}M^4
\nn\\
&&+ \frac{5}{2}k_5M^5\ln \frac{M}{\mu} + \frac{k_5+5k_6}{2}M^5
\label{chiral:expansion:sigma:term}
\\
&&+ 3k_7M^6\ln^2\frac{M}{\mu} + (k_7+3k_8)M^6\ln \frac{M}{\mu}+ \frac{k_8+6k_9}{2}M^6.
\nn
\eqa

\vspace{0.2 cm}
\noindent{\color{blue}\it Confronting ${\cal O}(q^6)$ $\chi$PT prediction with lattice QCD.} With the knowledge \eqref{mN:chiral:expansion:q6} at hand,
it is a good place to update the previous comparative analysis between the ${\cal O}(q^4)$ $\chi$PT and
lattice QCD predictions for the pion mass dependence of nucleon mass~\cite{Meissner:2005ba}.

We first specify the values of various input parameters.
To facilitate the numerical analysis, we set $\mu=m_N$.
We take the values of mesonic LECs from the latest lattice calculation by {\tt CLQCD} collaboration:
$l_3 = (2.2\pm0.9)\times10^{-3}$, $l_4 = (3.2\pm0.8)\times10^{-3}$~\cite{CLQCD:2023sdb}.
We then estimate the values of $M$ and $F$ from their physical values using the one-loop formula:
$M^2 = M_\pi^2 - M_\pi^4\ln(M_\pi/m_N)/(16\pi^2 F_\pi^2) - 2l_3\,M_\pi^4/F_\pi^2+\cdots$,
$F = F_\pi + M_\pi^2\ln(M_\pi/m_N)/(8\pi^2F_\pi) - l_4\,M_\pi^2/F_\pi+\cdots$, and obtain $M = 139.6 + 1.22 = 141$ MeV,
$F = 92.4 - 5.77 = 86.6$ MeV.
In the baryon sector, the axial-vector coupling in the chiral limit is taken to be ${\tt g_A}=1.13$~\cite{Yao:2017fym}, and
the value of $d_{16}$ is taken to be $-0.83$  $\text{GeV}^{-2}$~\cite{Yao:2017fym}.
We take the values of the LECs $c_{1\sim 4}$, $d_{18}$, $e_{14\sim 16}$, $\hat{e}_{20,36,38}$ from
Table~III Fit~I of \cite{Chen:2012nx}, which were determined by fitting the $\pi N$ scattering using the one-loop $\chi$PT prediction.

\begin{figure}[ht]
    \centering
    \includegraphics[width=0.5\textwidth]{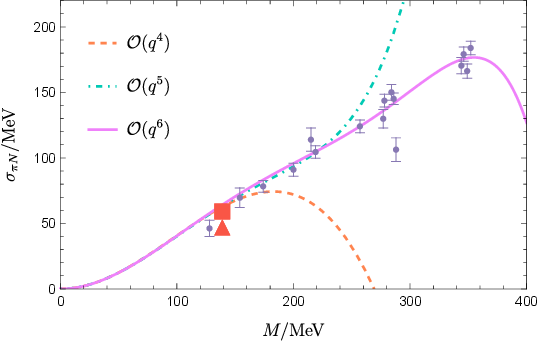}
    \caption{The $\pi N$ sigma term as a function of $M$.  The dashed, dot-dashed and solid curves represent the $\chi$PT predictions accurate at $\mathcal{O}(q^4)$,
    $\mathcal{O}(q^5)$ and  $\mathcal{O}(q^6)$, respectively. The lattice QCD predictions~\cite{Agadjanov:2023efe} are also juxtaposed for the sake of comparison.
    The solid triangle denotes the physical point value of $\sigma_{\pi N}$ determined by lattice approach, $46.9\pm 1.7$ MeV~\cite{Agadjanov:2023efe},
    and the solid square denotes that determined by Roy-Steiner equations, $59.1\pm 3.5$ MeV\cite{Hoferichter:2015dsa}.}
    \label{diag:sigma}
\end{figure}

To date our knowledge about the value of the tree-level contact term, especially $\hat{e}_1$ and $\hat{g}_1$, is rather limited.
Here we adopt a strategy to invoke the lattice data of the $\sigma_{\pi N}$ term to constrain their values.
First invert \eqref{mN:chiral:expansion:q6} to reexpress $m$ as a function of $m_N$, $M$ and various LECs,
then conduct a chiral expansion and truncate it through ${\cal O}(q^6)$.
Substituting this expression of $m$ to \eqref{chiral:expansion:sigma:term},
and taking the physical nucleon mass $m_N = \left(m_p + m_n\right)/{2} = 938.9$ MeV as input, we perform a weighted nonlinear least $\chi^2$ fit
for the latest lattice data about variation of $\sigma_{\pi N}$ with $M$~\cite{Agadjanov:2023efe}
to determine the values of these two unknown LECs.
The results are $\hat{e}_1 M^4 = 2.7 \pm 0.2$ MeV and $\hat{g}_1M^6 = 9.1\pm0.1$ MeV~\footnote{Note this value of $\hat{e}_1M^4$ is
consistent with the previous estimate based on the ${\cal O}(q^4)$ EOMS analysis of the $\sigma_{\pi N}$ term,  $2.3\pm 4$ MeV~\cite{Scherer:2012xha},
albeit with large errors. The value of $\hat{g}_1M^6$ presented here is new.}.
In Fig.~\ref{diag:sigma} we plot the $\chi$PT predictions for the $M$ dependence of $\sigma_{\pi N}$ in different chiral accuracy,
juxtaposed by the lattice data~\cite{Agadjanov:2023efe}.
From Fig.~\ref{diag:sigma} we observe that, the $\chi$PT predictions accurate at $\mathcal{O}(q^4)$, or even accurate at $\mathcal{O}(q^5)$,
can only describe the lattice data on the  $\sigma_{\pi N}$ in a rather limited range of pion mass ($M<200$ MeV).  In contrast,
the $\chi$PT prediction implementing $\mathcal{O}(q^6)$ correction can account for the
lattice data in a  broad range of $M$ ($M<350$ MeV).

\vspace{0.2 cm}

With the values of  $\hat{e}_1$ and $\hat{g}_1$ fixed, we can determine the nucleon mass in the chiral limit by inverting
\eqref{mN:chiral:expansion:q6} iteratively.
The solution is $m= 856.6\pm 1.7$ MeV, where the error originates from uncertainties of various LECs~\footnote{We caution the readers that the error estimation of $m$
might be overly optimistic, because the fitted values of $c_1$ through $c_4$ provided in \cite{Chen:2012nx},
which are most important LECs for our purpose, do not entail any uncertainty.}.
For the sake of clarity, we present \eqref{mN:chiral:expansion:q6} faithfully in a numerical format:
\bqa
m_N/\text{MeV} &= & \underbrace{856.6}_{m}+\underbrace{111}_{\mathcal{O}(q^2)} +\underbrace{(-14.4)}_{\mathcal{O}(q^3)}
\\
&+& \underbrace{(-9.40) + (- 4.48)}_{\mathcal{O}(q^4)}+\underbrace{(-4.02) + 4.42}_{\mathcal{O}(q^5)}
\nn\\
&+& \underbrace{0.775 + 1.97 +(- 2.39)}_{\mathcal{O}(q^6)}.
\nn
\label{nucleon:mass:chiral:limit}
\eqa
It is interesting to note that, the logarithmic and non-logarithmic terms in ${\cal O}(q^{5/6})$ corrections
accidentally cancel with each other to a large extent, so that the net ${\cal O}(q^{5/6})$ contributions are greatly suppressed.
Consequently, the EOMS prediction for the chiral expansion of nucleon mass exhibits a perfect convergence pattern around the physical point.

\begin{figure}[ht]
    \centering
    \includegraphics[width=0.5\textwidth]{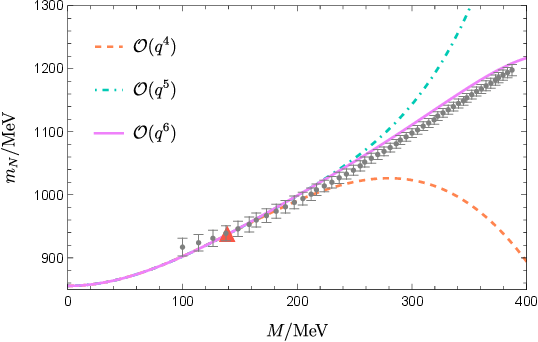}
    \caption{$M$ dependence of nucleon mass. The solid triangle corresponds to the physical point.
    The dashed, dot-dashed and solid curves represent the $\chi$PT predictions accurate at $\mathcal{O}(q^4)$,
    $\mathcal{O}(q^5)$ and  $\mathcal{O}(q^6)$, respectively. The lattice QCD predictions~\cite{Yang:2018nqn} are also juxtaposed for the sake of comparison.}
    \label{diag:mN}
\end{figure}

In light of \eqref{mN:chiral:expansion:q6}, with $m=856.6$ MeV as input, finally in Fig.~\ref{diag:mN} we plot $m_N$ as a function of $M$ at various level of chiral accuracy.
Similar to the pattern observed in Fig.~\ref{diag:sigma}, we observe that, the $\chi$PT predictions accurate at $\mathcal{O}(q^4)$, or even accurate at $\mathcal{O}(q^5)$,
tend to rapidly deviate from the recent {\tt $\chi$QCD} lattice prediction~\cite{Yang:2018nqn} when $M>200$ MeV.
On the other hand, once incorporating the $\mathcal{O}(q^6)$ correction, the $\chi$PT prediction can decently account for the
lattice data over a quite wide range of pion mass ($M<380$ MeV).

\vspace{0.2 cm}

\vspace{0.2 cm}
\noindent{\color{blue}\it Summary.} In this work we calculate the nucleon self energy to two-loop order
in the covariant $SU(2)$ $\chi$PT, and present the chiral expansion of nucleon mass to ${\cal O}(q^6)$ in EOMS scheme.
We have explicitly verified that the EOMS renormalization has passed highly nontrivial consistency test at two-loop order,
supported by the fact that all nonlocal divergences and PCB terms that are nonanalytic in $M_\pi$ have been exactly cancelled.
Taking the fitted values of the LECs determined by confronting the ${\cal O}(q^4)$ EOMS predictions with the $\pi N$ scattering data,
together with the aid of the latest lattice data on the $\sigma_{\pi N}$ term, we are able to determine the values of two unknown higher-order LECs $\hat{e}_1$ and $\hat{g}_1$.
We then present a state-of-the-art prediction for the value of the nucleon mass in the chiral limit: $856.6\pm 1.7$ MeV.
The EOMS scheme proves to possess a perfect convergence behavior for chiral expansion of nucleon mass around the physical point.
It is found that our ${\cal O}(q^6)$ prediction for the variation of nucleon mass with the pion mass
can decently account for the recent lattice results over a wide range of pion mass.

\begin{acknowledgments}
\noindent We thank Jing-Shu Dai for participating in the early stage of this work.
We are grateful to Li-Sheng Geng and Yi-Bo Yang for helpful discussions and comments about the manuscript.
We are also indebted to Yi-Bo Yang for providing us with the lattice data about pion mass dependence of nucleon mass in \cite{Yang:2018nqn}.
The work of L.-B. C. supported by the NNSFC Grant No.~12175048.
The work of S.~H., Y.~J. and Z.~M. is supported in part by the NNSFC Grant No.~11925506.
The work of Z.~M. is also supported in part by the NNSFC Grants No.~12347145 and No.~12022514.
\end{acknowledgments}

\begin{widetext}

\appendix

\section{Encountered counterterms in the EOMS scheme}
\label{counterterms:EOMS}

In this appendix, we show the explicit expressions for all the encountered counterterms.  
The subscript ``{\tt R}'' in all LECs has been suppressed for simplicity.

For purely mesonic sector, one has~\cite{Gasser:1983yg}
\beq
\delta_{\rm div}l_3 = \frac{1}{64\pi^2}\frac{1}{\epsilon},
\qquad    \delta_{\rm div}l_4 = \frac{-1}{16\pi^2}\frac{1}{\epsilon}.
\eeq

To facilitate the comparison with literature, we divide the counterterms in baryon sector into two categories. 
The first classes of counterterms,  labeled with a subscript $(1)$, are inserted in the one-loop counterterm diagrams, 
while the second classes of counterterms,  denoted by a subscript $(2)$,  serve to cancel the remaining UV divergences and PCB terms 
either from two-loop diagrams or one-loop diagrams of order $q^5$ and $q^6$.
The first classes are determined from the one-loop calculation of various observables in EOMS scheme, namely, 
the nucleon self energy~\cite{Fuchs:2003qc, Pascalutsa:2005nd}, axial or pion-nucleon form factors~\cite{Schindler:2006it, Ando:2006xy} and $\pi N$ 
scattering~\cite{Chen:2012nx, Siemens:2016hdi}. 
A comprehensive tabulation of EOMS counterterms is provided by Ref.~\cite{Chen:2012nx, Siemens:2016hdi}~\footnote{We remind the readers that 
Ref.~\cite{Chen:2012nx} contains some minor typos, which were later corrected by Ref.~\cite{Siemens:2016hdi}. Nevertheless, 
Ref.~\cite{Siemens:2016hdi} adopt a modified EOMS scheme, in which some finite PCP terms have also been included in the counterterms.}.

For $\delta_{\text{div}}$, we define
\begin{equation}
    \delta_{\text{div}} =
    \frac{1}{\epsilon}\frac{1}{16\pi^2F^2}\delta_{\text{div}, (1)} +
    \frac{1}{\epsilon^2}\frac{1}{256\pi^4F^4}\delta_{\text{div}, (2)}^{(2)} +
    \frac{1}{\epsilon}\frac{1}{256\pi^4F^4}\delta_{\text{div}, (2)}^{(1)} +
    \frac{1}{\epsilon}\frac{1}{256\pi^4F^4}\ln\frac{m}{\mu}\, \delta_{\text{div}, (2)}^{(1')}.
\end{equation}

The divergent counterterms inserted in the one-loop counterterm diagrams are as follows.
For nucleon mass and axial vector coupling in the chiral limit, we have
\begin{subequations}
\bqa
\delta_{\text{div}, (1)} m &=& -\frac{3m^3 {\tt g_A}^2}{2},\\
\delta_{\text{div}, (1)} {\tt g_A} &=& \frac{m^2 {\tt g_A}}{6}\big[12 - 6 {\tt g_A}^2-\left(3 c_2+8 c_3-40 c_4\right) m\big].
\eqa
\end{subequations}
For various LECs, we have  
\begin{subequations}
\begin{align}
    \delta_{\text{div}, (1)} c_1 &=
        -\frac{9}{2} c_1 m^2 {\tt g_A}^2+\frac{3 m {\tt g_A}^2}{8},\\
    \delta_{\text{div}, (1)} c_2 &=
        -\frac{1}{6} m^2 \big[\left(3 c_2+8 c_3+4 c_4\right) {\tt g_A}^2-c_4\big]-\frac{1}{2} m \left({\tt g_A}^2-1\right)^2,\\
    \delta_{\text{div}, (1)} c_3 &=
        -\frac{1}{12} m^2 \big[\left(21 c_2+54 c_3-52 c_4\right) {\tt g_A}^2+20 c_4\big]-\frac{1}{4} m \left({\tt g_A}^4-6 {\tt g_A}^2+1\right),\\
    \indent\\
    \delta_{\text{div}, (1)} \hat{d}_{16} &=
        \frac{1}{12} m {\tt g_A} \left(48 c_1 {\tt g_A}^2-72 c_1-c_2-16 c_3+32 c_4\right)+\frac{1}{2} {\tt g_A} \left(1-{\tt g_A}^2\right),\\
    \indent\nonumber\\
    \delta_{\text{div}, (1)} \hat{e}_1 &= -6 c_1 {\tt g_A}^2-6 c_1+\frac{3 c_2}{4}+3 c_3,\\
    \delta_{\text{div}, (1)}e_1 &=
        \frac{1}{192} \big[\left(25 c_2-8 c_3-20 c_4\right) {\tt g_A}^2+4 \left(2 c_2+12 c_3+5 c_4\right)\big]-\frac{\left({\tt g_A}^2-1\right)^2}{64 m},\\
    \delta_{\text{div}, (1)}e_{15} &= -\frac{1}{12} c_2 {\tt g_A}^2,\\
    \delta_{\text{div}, (1)}e_{16} &= 0,\\
    \delta_{\text{div}, (1)}\hat{e}_{20} &=
        -\frac{1}{4} c_1 {\tt g_A}^4+\left(\frac{c_1}{2}+\frac{c_2}{24}\right) {\tt g_A}^2-\frac{1}{4} \left(c_1+c_2\right),\\
    \delta_{\text{div}, (1)} \hat{e}_{36} &= \frac{{\tt g_A}^4}{4} c_1-\frac{{\tt g_A}^2}{24} \left(36 c_1+18 c_2+5 c_3-12 c_4\right) +\frac{\left({\tt g_A}^2-1\right)^2}{16 m}-\frac{1}{48} \left(36 c_1-5 c_2+18 c_3+24 c_4\right),\\
    \delta_{\text{div}, (1)} \hat{e}_{38} &=
        \frac{{\tt g_A}^2}{96} \left(144 c_1+39 c_2-8 c_3-4 c_4\right) +\frac{1}{24} \left(36 c_1-3 c_2-9 c_3+c_4\right) - \frac{\left({\tt g_A}^2-1\right)^2}{32 m}.
\end{align}
\end{subequations}

The $\delta_{\text{div}, (2)}$ terms are determined by the remaining UV divergences after summing all loop diagrams affiliated with nucleon self energy.
For the nucleon mass, we have
\begin{subequations}
\begin{align}
\delta_{\text{div}, (2)}^{(1)} m &=
    -\frac{3 m^5}{32}\bigg\{{\tt g_A}^2 \big[4 \left(c_2+2 c_3+13 c_4\right) m+6\big]-15 {\tt g_A}^4-20 c_4m-3\bigg\}\nn\\
    &-\frac{m^5}{1920} \bigg\{{\tt g_A}^2 \big[4 \left(231 c_2-670 c_3-775 c_4\right) m-210\big]+9105 {\tt g_A}^4-15\left(28 c_4 m+9\right)\bigg\},\\
\delta_{\text{div}, (2)}^{(1')} m &= \frac{3 m^5}{8} \bigg\{{\tt g_A}^2 \big[4 \left(c_2+2 c_3+13 c_4\right) m+6\big]-15 {\tt g_A}^4-20 c_4 m-3\bigg\}.
\end{align}
\end{subequations}
For the contact terms, we have
\begin{subequations}
\begin{align}
    \delta_{\text{div}, (2)}^{(2)} c_1 =&
    \frac{9 m^3 {\tt g_A}^4}{32} \left(25 c_1 m-6\right)+\frac{3m^3 {\tt g_A}^2}{32}  \big[12-30 c_1 m+4 \left(c_2+9 c_4\right) m\big])+\frac{3m^3}{32} \big[2 + 3 \left(5 c_1+4 c_4\right) m\big],\\
    \delta_{\text{div}, (2)}^{(1)} c_1 =&
    \frac{m^3 {\tt g_A}^2}{192} \big[513 c_1 m-4 \left(5 c_2+138 c_3+27 c_4\right) m-12 \left(192 \pi ^2 l_4+17\right)\big]\nn\\
    &+\frac{m^3 {\tt g_A}^4}{128}  \left(32-3755 c_1 m\right)+24 m^3 {\tt g_A}\pi^2 \hat{d}_{16} F^2 -\frac{m^3}{128}  \left(99 c_1 m+72 c_4 m+4\right),\\
    \delta_{\text{div}, (2)}^{(1')} c_1 =&
    \frac{3m^3 {\tt g_A}^4}{8} \left(14-75 c_1 m\right)+\frac{3m^3 {\tt g_A}^2}{4} \big[15 c_1 m-2 \left(c_2+9 c_4\right) m-2\big])-\frac{3m^3}{8} \big[3 \left(5 c_1+4 c_4\right) m+2\big],\\
    \delta_{\text{div}, (2)}^{(2)} \hat{e}_1 =&
    -\frac{7}{2} c_2 m^2 {\tt g_A}^2-18 c_3 m^2 {\tt g_A}^2+\frac{3}{4} c_4 m^2 \left(19 {\tt g_A}^2-3\right)+9 c_1 m^2 \left(9 {\tt g_A}^4-3 {\tt g_A}^2-1\right)\nn\\
    &-\frac{3}{16} m \left(17 {\tt g_A}^4-26 {\tt g_A}^2+3\right),\\
    \delta_{\text{div}, (2)}^{(1)} \hat{e}_1 =&
    96 \pi^2 m {\tt g_A}F^2\hat{d}_{16} \left(1-12 c_1 m\right)+\frac{3m {\tt g_A}^4}{32} \left(53-736 c_1 m\right)+\frac{3m}{32} \left(144 c_1 m+52 c_4 m+1\right)\nn\\
    &+\frac{3}{16} m {\tt g_A}^2 \big[6144 \pi ^2 c_1^2 F^2+256 \pi ^2 \left(l_3-l_4\right)-21\big]\nn\\
    &+\frac{1}{24} m^2 {\tt g_A}^2 \big[-9 \left(-16 c_3+7 c_4+192 \pi ^2 \hat{e}_1 F^2\right)+1152 c_1 \left(12 \pi ^2 l_4+1\right)+79 c_2\big],\\
    \delta_{\text{div}, (2)}^{(1')} \hat{e}_1 =&
    \frac{9m {\tt g_A}^4}{4} \left(3-88 c_1 m\right)-\frac{m {\tt g_A}^2}{4} \big[24 - \left(72 c_1+29 c_2+180 c_3-228 c_4\right) m\big]+\frac{9m}{4} \big[1+4\left(4 c_1+c_4\right) m\big]),\\
    \delta_{\text{div}, (2)}^{(2)} \hat{g}_1 =&
    -\frac{51{\tt g_A}^4}{4} c_1 +\frac{{\tt g_A}^2}{2} \left(33 c_1-2 c_2+9 c_3-3 c_4\right) -\frac{3}{4} \left(-3 c_1+c_2+4 c_3-2 c_4\right),\\
    \delta_{\text{div}, (2)}^{(1)} \hat{g}_1 =&
    384 \pi ^2 c_1 \hat{d}_{16} F^2 {\tt g_A}+\frac{3}{8} c_1 {\tt g_A}^4+c_1 \big[\frac{96 \pi ^2 c_2 F^2}{m}+192 \pi ^2 \left(l_3-l_4\right)-\frac{21}{8}\big]\nn\\
    &+{\tt g_A}^2 \bigg\{1536 \pi ^2 c_1^3 F^2+c_1 \big[192 \pi ^2 \left(l_3-l_4\right)+\frac{21}{4}\big]-\frac{67 c_2}{8}+24 \pi ^2 \hat{e}_1 F^2\bigg\}\nn\\
    &+24 \pi ^2 F^2 \left(c_2+4c_3\right) \left(l_4-2 l_3\right)-48 \pi ^2 \left(8 e_{14}+2 e_{15}+e_{16}+2 \hat{e}_{20}+4 \hat{e}_{36}+4 \hat{e}_{38}\right),\\
    \delta_{\text{div}, (2)}^{(1')} \hat{g}_1 =&
    \frac{35}{4} c_2 {\tt g_A}^2+6 c_1 \left({\tt g_A}^4-6 {\tt g_A}^2+1\right).
\end{align}
\end{subequations}

The finite PCB counterterms in EOMS scheme are decomposed into 
\begin{equation}
\begin{split}
    &\delta_{\text{eoms}} = \sum_{n}
    \frac{1}{16\pi^2F^2}\delta^{(n)}_{\text{eoms,(1)}}+
    \frac{1}{256\pi^4F^4}\delta^{(n)}_{\text{eoms,(2)}}\\
    &\delta^{(n)}_{\text{eoms,(i)}} = \delta^{(n, 0)}_{\text{eoms,(i)}} +
        \ln\frac{m}{\mu}\, \delta^{(n, 1)}_{\text{eoms,(i)}} +
        \ln^2\frac{m}{\mu} \, \delta^{(n, 2)}_{\text{eoms,(i)}}, ~ i=1 \text{~or~} 2,
\end{split}
\end{equation}
the superscript ${(n)}$ signifies the relative order between the $\widetilde{\text{MS}}$ renormalized quantity and the countered diagram.

Note we have retained the ${\cal O}(\epsilon)$ contribution in the finite EOMS counterterms.
For nucleon mass and axial coupling constant, we have
\begin{subequations}
\begin{align}
\frac{1}{m^3}\delta^{(2)}_{\text{eoms, (1)}} m =&
    -\frac{1}{8} \epsilon \left(6+\pi^2\right) {\tt g_A}^2
    + 3 {\tt g_A}^2\ln\frac{m}{\mu}
    -3 \epsilon {\tt g_A}^2 \ln^2\frac{m}{\mu},\\
\frac{1}{m^2}\delta^{(2)}_{\text{eoms},(1)} {\tt g_A} =&
        {\tt g_A}^3 - 2\left(2-{\tt g_A}^2\right) {\tt g_A}\ln\frac{m}{\mu}\nn\\
        &+\epsilon  \left(2 \left(2-{\tt g_A}^2\right) {\tt g_A}\ln^2\frac{m}{\mu}
        -2 {\tt g_A}^3\ln\frac{m}{\mu}+\frac{\left(6+\pi ^2\right) \left(2-{\tt g_A}^2\right) {\tt g_A}}{12}\right),\\
\frac{1}{m^2}\delta^{(3,0)}_{\text{eoms},(1)} {\tt g_A} =&
    \frac{{\tt g_A}}{36} \big[9 c_2+32 \left(c_3+c_4\right)\big]
    +\epsilon \frac{{\tt g_A}}{216}\big[9 \left(9+\pi ^2\right) c_2+8 \left(34+3 \pi ^2\right) c_3-8 \left(74+15 \pi ^2\right) c_4\big],\\
\frac{1}{m^2}\delta^{(3,1)}_{\text{eoms},(1)} {\tt g_A} =&
    \frac{{\tt g_A} }{3} \big[3 c_2+8 \left(c_3-5 c_4\right)\big]
    +\epsilon\frac{{\tt g_A}}{18} \big[9 c_2+32 \left(c_3+c_4\right)\big],\\
\frac{1}{m^2}\delta^{(3,1)}_{\text{eoms},(1)} {\tt g_A} =&
    -\epsilon\frac{ {\tt g_A}}{3} \big[3 c_2+8 \left(c_3-5 c_4\right)\big].
\end{align}
\end{subequations}

For various LECs, we have 
\begin{subequations}
\begin{align}
\frac{1}{m}\delta^{(1)}_{\text{eoms},(1)} c_1 =&
    \frac{3 {\tt g_A}^2}{8}-\frac{3}{4} {\tt g_A}^2 \ln\frac{m}{\mu }+\epsilon  \left(\frac{3}{4} {\tt g_A}^2 \ln ^2\frac{m}{\mu }-\frac{3}{4} {\tt g_A}^2 \ln
    \frac{m}{\mu }+\frac{\left(30+\pi ^2\right)}{32} {\tt g_A}^2\right),\\
\frac{1}{m^2}\delta^{(2)}_{\text{eoms},(1)} c_1 =&
    3 c_1 {\tt g_A}^2 + 9 c_1 {\tt g_A}^2 \ln\frac{m}{\mu}
    +\epsilon  \left(-9 c_1 {\tt g_A}^2 \ln ^2\frac{m}{\mu}-6 c_1 {\tt g_A}^2 \ln\frac{m}{\mu}-\frac{3\left(6+\pi ^2\right)}{8} c_1 {\tt g_A}^2\right),\\
\frac{1}{m}\delta^{(1)}_{\text{eoms},(1)} c_2 =&
        -\frac{1}{2} \left({\tt g_A}^4+2\right) + \left({\tt g_A}^2-1\right)^2 \ln\frac{m}{\mu }+\epsilon \big[\left(2+{\tt g_A}^4\right) \ln\frac{m}{\mu}-\left({\tt g_A}^2-1\right)^2 \ln ^2\frac{m}{\mu}\big],\\
\frac{1}{m^2}\delta^{(2)}_{\text{eoms},(1)} c_2 =&
    \frac{1}{9} \big[\left(9 c_2+16 c_3+14 c_4\right){\tt g_A}^2-2 c_4\big] + \frac{1}{3} \big[\left(3 c_2+8 c_3+4 c_4\right) {\tt g_A}^2-4 c_4\big] \ln\frac{m}{\mu}\nn\\
    & -\frac{1}{3} \epsilon \big[\left(3 c_2+8 c_3+4 c_4\right) {\tt g_A}^2-4 c_4\big] \ln^2\frac{m}{\mu}
    -\frac{2}{9} \epsilon \big[\left(9 c_2+16c_3+14 c_4\right) {\tt g_A}^2-2 c_4\big] \ln\frac{m}{\mu}\nn\\
    &-\frac{1}{216} \epsilon m\bigg\{\big[9 \left(6+\pi ^2\right) c_2+8 \left(22+3\pi ^2\right) c_3+4 \left(34+3 \pi ^2\right) c_4\big] {\tt g_A}^2-4 \left(10+3 \pi^2\right) c_4\bigg\},\\
\frac{1}{m}\delta^{(1)}_{\text{eoms},(1)} c_3 =&
    \frac{9 {\tt g_A}^4}{4} + \frac{1}{2} \left({\tt g_A}^4-6 {\tt g_A}^2+1\right) \ln\frac{m}{\mu} -\frac{1}{48} \epsilon \left(\left(\pi ^2-30\right) {\tt g_A}^4-6 \left(22+\pi ^2\right) {\tt g_A}^2+\pi^2+6\right)\nn\\
    &-\epsilon  \big[\frac{1}{2} \left({\tt g_A}^4-6 {\tt g_A}^2+1\right) \ln^2\frac{m}{\mu}+\frac{9}{2} {\tt g_A}^4 \ln\frac{m}{\mu }\big],\\
\frac{1}{m^2}\delta^{(2,0)}_{\text{eoms},(1)} c_3 =&
    -\frac{1}{8} c_2 {\tt g_A}^2+3 c_3 {\tt g_A}^2+\frac{2}{9} c_4 \left(1-17 {\tt g_A}^2\right)\nn\\
    &-\epsilon \bigg(\frac{1}{48} \left(45+7 \pi ^2\right) c_2 {\tt g_A}^2+\frac{3}{8} \left(6+\pi ^2\right)c_3 {\tt g_A}^2-\frac{1}{108} c_4 \big[\left(250+39 \pi ^2\right) {\tt g_A}^2-15 \pi^2-74\big]\bigg)\nn\\
\frac{1}{m^2}\delta^{(2,1)}_{\text{eoms},(1)} c_3 =&
    \frac{7}{2} c_2 {\tt g_A}^2+9 c_3 {\tt g_A}^2-\frac{2}{3} c_4 \left(13{\tt g_A}^2-5\right)
    +\epsilon \left(\frac{1}{4} c_2 {\tt g_A}^2-6 c_3 {\tt g_A}^2+\frac{4}{9} c_4 \left(17{\tt g_A}^2-1\right)\right) \nn\\
\frac{1}{m^2}\delta^{(2,2)}_{\text{eoms},(1)} c_3 =&
    \epsilon \left(-\frac{7}{2} c_2 {\tt g_A}^2-9 c_3 {\tt g_A}^2+\frac{2}{3} c_4 \left(13 {\tt g_A}^2-5\right)\right),\\
\frac{1}{m}\delta^{(1,0)}_{\text{eoms},(1)} \hat{d}_{16} =&
    \frac{1}{9} {\tt g_A} \left(-72 c_1 {\tt g_A}^2+72 c_1-2 c_2-17 c_3+19 c_4\right)\nn\\
    &+\epsilon {\tt g_A}\bigg(\frac{3\left(18+\pi^2\right){\tt g_A}^2-2\left(6+\pi ^2\right)}{6} c_1-\frac{82+3 \pi ^2}{432} c_2 -\frac{124+3 \pi^2}{27}  c_3 +\frac{2\left(73+3 \pi ^2\right)}{27}  c_4 \bigg),\\
\frac{1}{m}\delta^{(1,1)}_{\text{eoms},(1)} \hat{d}_{16} =&
    \frac{1}{6} {\tt g_A} \left(-48 c_1 {\tt g_A}^2+72 c_1+c_2+16 c_3-32 c_4\right)
    +\frac{2}{9} \epsilon {\tt g_A} \left(72 c_1 {\tt g_A}^2-72 c_1+2 c_2+17 c_3-19 c_4\right),\\
\frac{1}{m}\delta^{(1,2)}_{\text{eoms},(1)} \hat{d}_{16} =&
    \frac{1}{6} \epsilon {\tt g_A} \left(48 c_1 {\tt g_A}^2-72 c_1-c_2-16 c_3+32 c_4\right)
\end{align}
\end{subequations}

Additional finite counterterms of nucleon mass reads
\begin{subequations}
\begin{align}
\frac{1}{m^5}\delta^{(4,0)}_{\text{eoms},(2)} m =&
    \frac{\left(344 \pi ^2-145\right)}{512}  {\tt g_A}^4+\frac{\left(136 \pi ^2-575\right)}{256}  {\tt g_A}^2+\frac{\left(120 \pi ^2-97\right)}{512}\\
\frac{1}{m^5}\delta^{(4,1)}_{\text{eoms},(2)} m =&
        \frac{1}{32} \left(607 {\tt g_A}^4-14 {\tt g_A}^2-9\right)\\
\frac{1}{m^5}\delta^{(4,2)}_{\text{eoms},(2)} m =&
        \frac{9}{4} \left(5 {\tt g_A}^4-2 {\tt g_A}^2+1\right)\\
\frac{1}{m^6}\delta^{(5,0)}_{\text{eoms},(2)} m =&
        \frac{{\tt g_A}^2}{28800}\bigg\{27 \left(200 \pi ^2-1433\right) c_2+25 \big[\left(816 \pi ^2-2546\right) c_3+\left(1464 \pi ^2-11177\right) c_4\big]\bigg\}\nn\\
        &+\frac{1}{128} \left(200 \pi^2-191\right) c_4\\
\frac{1}{m^6}\delta^{(5,1)}_{\text{eoms},(2)} m =&
        \frac{1}{120}\big[\left(231 c_2-670 c_3-775 c_4\right) {\tt g_A}^2-105 c_4\big]\\
\frac{1}{m^6}\delta^{(5,2)}_{\text{eoms},(2)} m =&
        3 \big[5 c_4-\left(c_2+2 c_3+13 c_4\right) {\tt g_A}^2\big],\\
\end{align}
\end{subequations}

Additional finite counterterms of the contact terms read
\begin{subequations}
\begin{align}
\frac{1}{m^3}\delta^{(3, 0)}_{\text{eoms, (2)}} c_1 =&
    -\frac{1}{64} \big[\left(46 \pi ^2-26\right) {\tt g_A}^4+\left(118-20 \pi ^2\right) {\tt g_A}^2-10 \pi ^2+7\big]\\
\frac{1}{m^3}\delta^{(3, 1)}_{\text{eoms, (2)}} c_1 =&
    \frac{1}{8} \left(-384 \pi ^2 \hat{d}_{16} F^2 {\tt g_A}+2 \left(96 \pi ^2 l_4+17\right) {\tt g_A}^2-8 {\tt g_A}^4+1\right)\\
\frac{1}{m^3}\delta^{(3, 2)}_{\text{eoms, (2)}} c_1 =&
    \frac{3}{2} \left(1-6 {\tt g_A}^4\right)\\
\frac{1}{m^4}\delta^{(4, 0)}_{\text{eoms, (2)}} c_1 =&
    \frac{1}{512} \big[43 \left(209+40 \pi ^2\right) {\tt g_A}^4+2 \left(552 \pi ^2-2651\right) {\tt g_A}^2+600 \pi ^2-629\big]c_1\nn\\
    &+\frac{\left(613+168 \pi ^2\right)}{1152} c_2 {\tt g_A}^2+\frac{\left(8 \pi ^2-95\right)}{16}  c_3 {\tt g_A}^2+\frac{1}{32} c_4 \big[\left(85+42 \pi ^2\right) {\tt g_A}^2+30 \pi^2-31\big]\\
\frac{1}{m^4}\delta^{(4, 1)}_{\text{eoms, (2)}} c_1 =&
    \frac{1}{32} c_1 \left(3755 {\tt g_A}^4-342 {\tt g_A}^2+99\right)+\frac{5}{12} c_2 {\tt g_A}^2+\frac{23}{2} c_3  {\tt g_A}^2+\frac{9}{4} c_4 \left({\tt g_A}^2+1\right)\\
\frac{1}{m^4}\delta^{(4, 2)}_{\text{eoms, (2)}} c_1 =&
    \frac{45}{4} c_1 \left(5 {\tt g_A}^4-2 {\tt g_A}^2+1\right)+3 c_2 {\tt g_A}^2+9 c_4 \left(3 {\tt g_A}^2+1\right)
\end{align}
\end{subequations}

\begin{subequations}
\begin{align}
\frac{1}{m}\delta^{(1,0)}_{\text{eoms},(2)} \hat{e}_{1} =&
    {\tt g_A}^2 \big[\frac{39}{32}-\pi^2 (1920 c_1^2 F^2-48 l_3+48 l_4-1)\big]+96 \pi^2 \hat{d}_{16} F^2 {\tt g_A}\nn\\
    &-\frac{3}{64} {\tt g_A}^4 \big[-48 \zeta_3+305+\pi^2 (32 \ln 2-50)\big]-\frac{3}{64} \left(41+10 \pi^2\right),\\
\frac{1}{m}\delta^{(1,1)}_{\text{eoms},(2)} \hat{e}_{1} =&
    -2304 \pi ^2 c_1^2 F^2 {\tt g_A}^2-\frac{3}{8} m \left(512 \pi^2 \hat{d}_{16} F^2 {\tt g_A}+\big[256 \pi^2 \left(l_3-l_4\right)-78\big] {\tt g_A}^2+69{\tt g_A}^4+1\right),\\
\frac{1}{m}\delta^{(1,2)}_{\text{eoms},(2)} \hat{e}_{1} =&
    -\frac{3}{2} \left(5 {\tt g_A}^4+{\tt g_A}^2+3\right),\\
\frac{1}{m^2}\delta^{(2,0)}_{\text{eoms},(2)} \hat{e}_{1} =&
    \frac{1}{4} c_1 \big(3072 \pi^2 \hat{d}_{16} F^2 {\tt g_A}+\big[46-\pi ^2 \left(1536 l_4+35\right)\big] {\tt g_A}^2+\left(129 \pi ^2-140\right) {\tt g_A}^4-30 \pi ^2+13\big)\nn\\
    &-\frac{1}{288}\left(874+159 \pi ^2\right) c_2 {\tt g_A}^2-\frac{1}{8} \left(261+31 \pi ^2\right) c_3 {\tt g_A}^2\nn\\
    &+\frac{3}{16} c_4 \big[\left(293+26 \pi ^2\right) {\tt g_A}^2-10 \pi ^2+3\big]+48 \pi^2 \hat{e}_1 F^2 {\tt g_A}^2,\\
\frac{1}{m^2}\delta^{(2,1)}_{\text{eoms},(2)} \hat{e}_{1} =&
    6 c_1 \left(384 \pi ^2 \hat{d}_{16} F^2 {\tt g_A}-2 \left(96 \pi ^2 l_4+11\right) {\tt g_A}^2+32 {\tt g_A}^4-9\right)\nn\\
    &-\frac{26}{3} c_2 {\tt g_A}^2-6 c_3 {\tt g_A}^2+\frac{3}{2} c_4 \left(7{\tt g_A}^2-13\right)+144 \pi ^2 \hat{e}_1 F^2 {\tt g_A}^2,\\
\frac{1}{m^2}\delta^{(2,2)}_{\text{eoms},(2)} \hat{e}_{1} =&
    -\frac{31}{4} c_2 {\tt g_A}^2-63 c_3 {\tt g_A}^2+6 c_4 \left(19 {\tt g_A}^2-3\right)+18 c_1 \left(15 {\tt g_A}^4+3 {\tt g_A}^2-4\right).
\end{align}
\end{subequations}

\section{Analytical expression of the two-loop self-energy}
\label{analytical:expression:two:loop:Sigma}

In this section, we present the chiral-expanded two-loop self energy. 
As mentioned in the main text, we need to analytically calculate $\tilde\Sigma^{(2)}(\tilde{m}, \tilde{m})$, and expand it around $m$.
For those diagrams of nominally $\mathcal{O}(q^6)$, we simply replace $\tilde{m}$ by $m$. 
For those diagrams of nominally $\mathcal{O}(q^5)$, we replace $\tilde{m}$ by $m - 4c_1M^2$.

After chiral expansion,  the two-loop nucleon self energy can be decomposed into
\bqa
\tilde\Sigma^{(2)}(\tilde{m}, \tilde{m}) &=&\frac{i}{2^{13}\pi^4 F^4}\left(\frac{\mu^2}{m^2}\right)^{2\epsilon}\bigg[\frac{\Sigma^{(2)}_{-2}}{\epsilon^2}+\frac{\Sigma^{(2)}_{-1}}{\epsilon}
+ \Sigma^{(2) \texttt{PCB}}_{0}+\Sigma^{(2) \texttt{PCP}}_{0}\bigg].
\label{two:loop:self:energy:analytic}
\eqa
where the finite part has been divided into the PCB and PCP pieces. 
For clarity, we introduce the symbol $\lambda$ as a bookkeeping device to signify 
the nominal chiral order of each two-loop diagrams, which should be set to 1 in the end.

The coefficient of the double pole reads
\bqa
\Sigma^{(2)}_{-2} &=&-\lambda^5 3m \bigg\{m^4\left[89 {\tt g_A}^4-58 {\tt g_A}^2-3\right] + 4 m^2 M^2 \left[19 {\tt g_A}^4-20 {\tt g_A}^2+2\right] +  M^4\left[22 {\tt g_A}^4-52 {\tt g_A}^2+6\right]\bigg\}\nonumber\\
&+&\lambda^6  \bigg\{m^6 \left[\left(-60 c_2-152   c_3+484 c_4\right) {\tt g_A}^2+60 c_4\right]\nonumber\\&+& m^4 M^2 \left[5340 c_1 {\tt g_A}^4-8 \left(387 c_1+14 c_2+48 c_3-90 c_4\right) {\tt g_A}^2-36 \left(5 c_1+4 c_4\right)\right]\nonumber\\&+& m^2 M^4 \left[2736 c_1 {\tt g_A}^4-24 \left(68 c_1+8 c_2+24 c_3-19 c_4\right) {\tt g_A}^2+72 \left(4 c_1-c_4\right)\right]\nonumber\\
&+&M^6 \left[264 c_1 {\tt g_A}^4-6 \left(48 c_1+15 c_2+12 c_3-8 c_4\right) {\tt g_A}^2+24 \left(-3 c_1+c_2+4 c_3-2 c_4\right)\right] \bigg\},
\eqa
and the coefficient of the single pole reads
\bqa
\Sigma^{(2)}_{-1} &=& \frac{\lambda^5}{4m}\bigg\{m^6
\big[{\tt g_A}^4 353+14{\tt g_A}^2 +9\big]
\nn
\\
&+& m^4 M^2 \big[-32 {\tt g_A}^4 (27 \ln\frac{M}{m}+22)+1312{\tt g_A}^2+16\big]
\nn\\
&+& m^3 M^3\pi  {\tt g_A}^2 \big[816 {\tt g_A}^2-768 \big]
\nn\\
&+&m^2 M^4 \big[ {\tt g_A}^4(96 \ln\frac{M}{m}-1476)+24 {\tt g_A}^2 (-64 \ln\frac{M}{m}+79)+288 \ln\frac{M}{m}-132\big]
\nn\\
&+&m M^5\pi{\tt g_A}^2 \big[606   {\tt g_A}^2-816  \big]+M^6\big[{\tt g_A}^4
(528 \ln\frac{M}{m}-640)+{\tt g_A}^2 (848-960 \ln\frac{M}{m})\big]  \bigg\}
\nn\\
&+&\frac{\lambda^6}{4}\bigg\{ m^6 \big[28 c_4-\frac{4}{15} \left(591 c_2+610 c_3+505 c_4\right) {\tt g_A}^2\big]
\nn\\
&+&m^4 M^2 \big[-24148 c_1 {\tt g_A}^4+\frac{8}{3} \left(4095 c_1-4 \left(38 c_2+78 c_3-387 c_4\right)\right) {\tt g_A}^2+36 \left(11 c_1+8 c_4\right)\big]
\nn\\
&+& 64 m^3 M^3 {\tt g_A}^2\pi \big[3 c_ 2+8 (c_ 3-5 c_ 4)\big] 
\nn\\
&+& m^2 M^4 \big[384 c_ 1 {\tt g_A}^4 (27 \ln\frac{M}{m}-25)-8{\tt g_A}^2 (864 c_ 1 \ln\frac{M}{m}+c_ 2 (-204 \ln\frac{M}{m}+67)\nonumber\\
&+&  c_ 3(-528 \ln\frac{M}{m}+168)+ c_ 4 (720 \ln\frac{M}{m}-654))+ 48 c_ 4 (24\ln\frac{M}{m}-11)-1728 c_ 1\big]\nonumber\\
&+&m M^5\pi{\tt g_A}^2 \big[8   (576 c_ 1+5 c_ 2+24 (5 c_ 3-9
   c_ 4)) -6528 c_ 1 {\tt g_A}^2\big]
\nn\\
&+&M^6 \big[48 c_ 1 {\tt g_A}^4 (-8\ln\frac{M}{m}+43)+{\tt g_A}^2 (96 c_ 1 (72 \ln\frac{M}{m}+5)-4 c_ 2 (48 \ln\frac{M}{m}+311)
\nn\\
&-& \frac{64}{3} (c_ 3 (6 \ln\frac{M}{m}+73)+2 c_ 4 (66 \ln\frac{M}{m}-43)))
\nn\\
&+& 48  c_ 1(24\ln\frac{M}{m}+19)-384c_2\ln\frac{M}{m}-192c_3(8\ln\frac{M}{m}+1)+768c_4\ln\frac{M}{m}\big] \bigg\}.
\eqa

The finite PCB and PCP terms in \eqref{two:loop:self:energy:analytic} are
\begin{subequations}
\bqa
\Sigma^{(2) \texttt{PCB}}_{0} &=& \frac{\lambda^5}{144m}\bigg\{m^6 \big[  -9 (5137+488 \pi ^2) {\tt g_A}^4+18 (961+392 \pi ^2) {\tt g_A}^2+9
(120 \pi ^2-97)  \big]
\\
&+& m^4 M^2 \big[  288 {\tt g_A}^4 (108 \ln^2\frac{M}{m}+72 \ln\frac{M}{m}+13
\pi ^2-488)+576 (227+2 \pi ^2) {\tt g_A}^2\nonumber\\
&-&288 (10 \pi ^2-7)  \big]\nonumber\\
&+&m^3 M^3\pi  {\tt g_A}^2 \big[  27648  (2 \ln \frac{M}{m}-1+\ln4)-1728 {\tt
g_A}^2 (34 \ln \frac{M}{m}-3+17 \ln4)  \big]\nonumber\\
&+& m^2 M^4 \big[ -72 {\tt g_A}^4 (48 \ln ^2\frac{M}{m}+1608 \ln \frac{M}{m}+3 (48
\zeta (3)-443)+\pi ^2 (190-96 \ln 2))\nonumber\\
&+& 144 {\tt g_A}^2 (384 \ln ^2(\frac{M}{m})+312 \ln \frac{M}{m}+2 \pi ^2+75) \nonumber\\
&+& 216   (-48 \ln ^2\frac{M}{m}+40 \ln (\frac{M}{m})+6 \pi ^2-25)  \big]  \bigg\}\nonumber\\
&+& \frac{\lambda^6}{144} \bigg\{  m^6 \big[ (-\frac{36}{25} (10299+200 \pi ^2) c_ 2+8 (24 \pi ^2-4601) c_ 3+52
(755+408 \pi ^2) c_ 4) {\tt g_A}^2 \nonumber\\
&+& 36 (200 \pi ^2-191) c_ 4  \big]\nonumber\\
&+&  m^4 M^2 \big[  36 (31333+2440 \pi ^2) c_ 1 {\tt g_A}^4+8 (-9 (4261+1704 \pi ^2) c_ 1-2 (2633+264 \pi ^2) c_ 2 \nonumber\\
&-&  144 (97+16 \pi ^2) c_ 3+72 (683+6 \pi ^2)
c_ 4) {\tt g_A}^2-36 ((600 \pi ^2-629) c_ 1+16 (30 \pi ^2-31) c_ 4) \big] \nonumber\\
&-& 384 m^3 M^3 \pi {\tt g_A}^2   \big[ 9 c_ 2 (4 \ln \frac{M}{m}-3+\ln 16)+16 (c_3 (6 \ln \frac{M}{m}-5+\ln 64)\nonumber\\
&+& c_ 4 (-30 \ln\frac{M}{m}+13-15\ln 4))   \big] \nonumber\\
&+& m^2 M^4 \big[   144 ({\tt g_A}^2 (-(c_ 2 (408 \ln ^2\frac{M}{m}+36 \ln\frac{M}{m}+24 \pi ^2+133)\nonumber\\
&+& 4 c_ 3 (264 \ln ^2\frac{M}{m}+168 \ln\frac{M}{m}+8 \pi ^2+303)))\nonumber\\
&-& 2 c_ 4 ({\tt g_A}^2 (-720 \ln ^2\frac{M}{m}-696 \ln \frac{M}{m}+2 \pi
^2+309)+3 (48 \ln ^2\frac{M}{m}+8 \ln \frac{M}{m}-6 \pi ^2+25))\nonumber\\
&-&8c_ 1 ({\tt g_A}^4 (324 \ln^2\frac{M}{m}+432 \ln \frac{M}{m}+39 \pi ^2-1184)\nonumber\\
&+& 2 {\tt g_A}^2 (-108 \ln ^2\frac{M}{m}-72 \ln\frac{M}{m}+\pi ^2+182)-30\pi ^2+13))   \big] \nonumber\\
&+& 48  m M^5 \pi  {\tt g_A}^2 \big[ -288 c_ 1 (4 (6 \ln \frac{M}{m}+1+\ln 64)-{\tt g_A}^2 (34 \ln \frac{M}{m}+14+17 \ln 4)) \nonumber\\
&-& 5 c_ 2 (12 \ln \frac{M}{m}-37+6 \ln4)-144 c_ 3 (10 \ln \frac{M}{m}-13+5 \ln 4)\nonumber\\
&+& 432 c_ 4 (6 \ln \frac{M}{m}-7+\ln 64)     \big] \bigg\}.
\nn\\
\Sigma^{(2) \texttt{PCP}}_{0} &=& \frac{\lambda^5}{144m}\bigg\{
m M^5 \pi  {\tt g_A}^2  \big[  1728 (40 \ln \frac{M}{m}-35+17 \ln4)-72 {\tt g_A}^2 (942 \ln \frac{M}{m}-227+303 \ln4)  \big]\nonumber\\
&+& M^6 \big[ 96 {\tt  g_A}^4 (126 \ln ^2\frac{M}{m}-738 \ln\frac{M}{m}+84 \pi
^2+1241) \nonumber\\
&+& 192 {\tt  g_A}^2 (252 \ln ^2\frac{M}{m}+72 \ln\frac{M}{m}+96 \pi ^2-509)+576 (12 \ln
\frac{M}{m}-5)  \big]\bigg\}\nonumber\\
&+& \frac{\lambda^6}{144} \bigg\{ M^6 \big[  288 c_ 1 {\tt g_A}^4 (48 \ln ^2\frac{M}{m}+1704 \ln
\frac{M}{m}+144 \zeta_3-5889 \nonumber\\
&+& \pi ^2 (190-96 \ln 2))+8 {\tt g_A}^2 (216 c_ 1 (-288 \ln ^2\frac{M}{m}-376 \ln \frac{M}{m}+4 \pi
^2+759)\nonumber\\
&+&c_ 2 (4752 \ln ^2\frac{M}{m}+4464 \ln \frac{M}{m}-558
\pi ^2-16347)\nonumber\\
&+& 8 c_ 3 (2016 \ln ^2\frac{M}{m}+1248 \ln\frac{M}{m}-411 \pi ^2+890) 
\nn\\
&+& 16 c_ 4 (792 \ln^2\frac{M}{m}+156 \ln \frac{M}{m}+357 \pi ^2-1541))
\\
&+& 96 (-9 c_1 (144 \ln ^2\frac{M}{m}+8 \ln \frac{M}{m}+10 \pi ^2+47)+6 c_ 2(48 \ln ^2\frac{M}{m}+\pi^2+6) 
\nn\\
&+& 24 c_ 3 (48 \ln ^2\frac{M}{m}+12 \ln \frac{M}{m}+\pi^2+6)
+4 c_ 4 (-72 \ln ^2\frac{M}{m}+84 \ln \frac{M}{m}+9 \pi^2-23)) \big] \bigg\}.
\nn
\eqa
\end{subequations}

\end{widetext}


\end{document}